# Mean Field Theory for the Quantum Rabi Model, Inconsistency to the Rotating Wave Approximation

*Ghasem Asadi Cordshooli, Mehdi Mirzaee\**
*\* m-mirzaee@araku.ac.ir*
*Physics of Department, faculty of Science, Arak University, Arak  38156-8-8349, Iran*

*Abstract*

Time evolution of pertinent operators in the Rabi Hamiltonian and its rotating wave approximation (RWA) version, the Jaynes-Cummings model (JCM), in the Heisenberg picture, gives systems of nonlinear differential equations (NDEs). Considering well localized atom, the mean field theory (MFT) was applied to replace the operators by equivalent expectation values. The Rabi model was reduced to a fourth orders NDE describing atoms position. Solution by the harmonic balance method (HBM) showed good accuracy and consistency to the numerical results, which introduces it as a useful tool in the quantum dynamics studies. The NDEs describing the JCM in the Heisenberg picture structurally prevent applying the MFT and shows inconsistency to the Ehrenfest's theorem, contrary to the Rabi model.

## 1. Introduction

Interaction of a two level atom by a radiation field was introduced by I. I. Rabi [1] in 1937. Applying the RWA [2, 3] to the Rabi model, i.e. discarding the counter rotating terms [4], results in the JCM [5]. The JCM as the RWA version [6] of the Rabi model is of great interest in the quantum information theory [7-9], and has applied widely in the theory of laser cooling and trapped ion dynamics [10-16]. Also, the Rabi model has many important applications in the quantum batteries [17], trapped ions [18], quantum coherent heat engine [19], cavity cooling [20], quantum dots [21] and superfluid Bose-Einstein condensate [22].

Whilst the RWA has been considered as a utility to provide simple framework for theoretical study of the atom-photon interaction in different systems, it tends to wrong results in a variety of situations. For instance applying adiabatic elimination after the RWA will eliminates some necessary terms in the two-photon-atom coupling [23]. It was also revealed that applying the RWA in the atom-cavity system in the absence of external driving, relates to the entropy operator and generates irreversible time evolution, as for the coupled resonator [24]. The output fundamental field component of a two level atom in a ring cavity beyond the RWA resulted in the closed loop and usual bistability [25]. It was also shown that the equation of motion derived by the Hamiltonian under the RWA is inconsistent with the Ehrenfest theorem [26]. According to another report, the RWA is invalid in the strict weak interaction and so it doesn't gives valid low temperature dynamics [27]. In the coherent control of a driven three level system, counter rotating terms have great influence on the evolution of atomic population and spontaneous emission spectrum [28].

The Rabi model has been solved by different methods [29-31] and its full eigenvalues spectrum obtained [32]. The model has also been considered from the integrability viewpoints [33-36]. In this paper the time evolutions of the operators contributed in the Rabi Hamiltonian will be calculated in the Heisenberg picture. The operators will be replaced by their expectation values using the MFT [37] for a well localized atom to reduce the system of NDEs to a higher order NDE that will be solved by the HBM [38]. The error analysis of the solution will be given for different orders of the HBM in addition to comparing to the numerical results for typical parameter values. At last effects of the RWA will be considered.

The paper is structured as follows. In the next section, the Rabi and JCM Hamiltonians are briefly introduced and their time evolutions of the constituent operators are calculated. The resulted system of NDEs for the Rabi model is solved in the section 3. The numerical and error considerations for the Rabi model and the effects of the RWA are also given in this section. The section 4 devotes to a discussion about the results and their relations to the multiparticle versions of the models, namely the Dicke and Tavis-Cummings models. A summary of the conclusions are given in the last section.

## 2. Model and Time Evolution

The Rabi model was proposed to describe the interaction of the classical magnetic field with the nucleus spin. From an abstract viewpoint, this model describes the interaction of one mode of a field with a two-state system, for instance a single mode of electric field interacting with a two level atom. Fully quantized version of this model is given by

$$\widehat{H}_R = \frac{\hbar\omega}{2}\widehat{\sigma}_z + \hbar\nu\hat{a}^\dagger\hat{a} + \hbar\lambda\widehat{\sigma}_x(\hat{a} + \hat{a}^\dagger). \tag{1}$$

The term $\frac{\hbar\omega}{2}\widehat{\sigma}_z$ shows the Hamiltonian of a two level atom with $\widehat{\sigma}_z = |e\rangle\langle e| - |g\rangle\langle g|$ and $\omega$ as the atomic transient frequency. The $\hbar\nu\hat{a}^\dagger\hat{a}$ describes a cavity mode by the frequency $\nu$ and $\hbar\lambda\widehat{\sigma}_x(\hat{a} + \hat{a}^\dagger)$ is the field-dipole interaction Hamiltonian.

The annihilation operator in the Hamiltonian (1), may be replaced by, $\hat{a}(t) = (\frac{m\nu}{2\hbar})^{1/2}(\hat{x}(t) + \frac{i}{m\nu}\hat{p}(t))$, [39] in which the canonical position and momentum operators, $\hat{x}(t)$ and $\hat{p}(t) = \dot{\hat{x}}(t)$, are field quadrature operators, equivalent to the classical $x(t)$ and $\dot{x}(t) = p(t)$ functions appeared in the electric and magnetic fields satisfying the Maxwell's equations. These operators are physical observables [3] satisfying the position-momentum commutation relation $[\hat{x}(t), \hat{p}(t)] = i\hbar$. The result is

$$\widehat{H}_R = \frac{\hbar\omega}{2}\widehat{\sigma}_z + \frac{1}{2m}\hat{p}^2(t) + \frac{1}{2}m\nu^2\hat{x}^2(t) + \hbar k\widehat{\sigma}_x(t)\hat{x}(t). \tag{2}$$

In this paper, the Hamiltonian (2) describes a two level ion in a harmonic trap interacting with a laser field. The second and third terms define the kinetic and potential energies of the ion, respectively. The last term shows the ion's electric dipole interacting with quantized laser field written in terms of the field quadrature operator $\hat{x}(t)$. In this way, $k = \lambda(2m\nu/\hbar)^{1/2}$ wherein $\lambda = -(\hbar\nu/\epsilon_0 V)^{1/2}\sin(kz)$ shows the atom-field interaction strength [4]. Applying the Heisenberg equation [4],

$$\frac{d\widehat{O}}{dt} = \frac{i}{\hbar}[\widehat{H}, \widehat{O}], \tag{3}$$

for the operators $\widehat{\sigma}_x, \widehat{\sigma}_y, \widehat{\sigma}_z, \hat{x}$ and $\hat{p}$ with the Hamiltonian (2), tends to following set of first order NDEs

$$\frac{d\widehat{\sigma}_x(t)}{dt} = -\omega\widehat{\sigma}_y(t), \tag{4}$$

$$\frac{d\widehat{\sigma}_y(t)}{dt} = \omega\widehat{\sigma}_x(t) - 2k\widehat{\sigma}_z(t)\hat{x}(t), \tag{5}$$

$$\frac{d\widehat{\sigma}_z(t)}{dt} = 2k\widehat{\sigma}_y(t)\hat{x}(t), \tag{6}$$

$$\frac{d\hat{x}(t)}{dt} = \frac{1}{m}\hat{p}(t), \tag{7}$$

$$\frac{d\hat{p}(t)}{dt} = -m\nu^2\hat{x}(t) - \hbar k\widehat{\sigma}_x(t). \tag{8}$$

In the next section we try to solve the system of NDEs (4) to (8) by considering $|g, \alpha\rangle$ as the initial state of the system.

## 3. Solution

### 3.1. The Mean Field Theory

In order to solve the systems of NDEs (4) to (8) all pertinent terms may be replaced by their expectation values. The nonlinear terms appeared in the NDEs (5) and (6) are in the form of $\hat{x}(t)\hat{\sigma}_x$ and $\hat{x}(t)\hat{\sigma}_z$ for which it is possible to use the quantum MFT [37] if we consider a well localized ion. The position operator may be written as $\hat{x} = \langle\hat{x}\rangle + \delta\hat{x}$ which $\langle\hat{x}\rangle$ shows the expectation value of the operator and $\delta\hat{x}$ holds for its fluctuations around the mean value. For a well localized atom, $\delta\hat{x} \approx 0$ and we can use $\hat{x} \approx \langle\hat{x}\rangle$. In this way the product nonlinear terms including the operator $\hat{x}$ may be considered as

$$\hat{x}\hat{\sigma}_i = \langle\hat{x}\rangle\hat{\sigma}_i + (\hat{x} - \langle\hat{x}\rangle)(\langle\hat{\sigma}_i\rangle + \delta\hat{\sigma}_i)$$
$$\approx \langle\hat{x}\rangle\hat{\sigma}_i + \hat{x}\langle\hat{\sigma}_i\rangle - \langle\hat{x}\rangle\langle\hat{\sigma}_i\rangle. \tag{9}$$

Calculating the expectation values of both sides of the equation (9) gives

$$\langle \hat{x}\hat{\sigma}_i \rangle \approx \langle \hat{x} \rangle \langle \hat{\sigma}_i \rangle. \tag{10}$$

Whilst we described the MFT for the position operator, it is equivalent to the field operator as both are proportional to $\hat{a} + \hat{a}^\dagger$. Representing the expectation values of all operators as $\langle \hat{O} \rangle = O$, hereafter we can replace the expectation values of the NDEs by removing the hat sign from the operators. To convert the system of NDEs (4) to (8) to a higher order NDE, the equations (4) and (6) may be combined and integrated to obtain

$$\sigma_z(t) = -1 - \frac{2k}{\omega} \int_0^t \frac{d\sigma_x(t)}{dt} x(t)\, dt. \tag{11}$$

The constant value in the right hand side of (11) is the expectation value of the operator $\hat{\sigma}_z(t)$ at $t = 0$ wherein the atom considered in the ground state, i.e. $\sigma_z(0) = \langle g|\hat{\sigma}_z(t)|g\rangle = -1$. Using $\sigma_y(t)$ from (4) and $\sigma_z(t)$ given by (11), the equation (5) converts to

$$\frac{d^2\sigma_x(t)}{dt^2} + \omega^2 \sigma_x(t) + 2k\lambda\omega x(t) + 4k^2\lambda^2 x(t) \int_0^t \frac{d\sigma_x(t)}{dt} x(t)\, dt = 0. \tag{12}$$

The equation (12) may be written in terms of the function $x(t)$ and its derivatives. Toward this end, one can remove $p(t)$ from the equations (7) and (8) to obtain the following result for $\sigma_x(t)$.

$$\sigma_x(t) = -\frac{k}{2\lambda\nu}\left(\frac{d^2 x(t)}{dt^2} + \nu^2 x(t)\right) \tag{13}$$

Using (13) and its derivatives, the equation (12) takes the form

$$\frac{d^4 x(t)}{dt^4} + (\omega^2 + \nu^2)\frac{d^2 x(t)}{dt^2} + \omega^2\nu^2 x(t) - 4\lambda^2\nu\omega x(t) + 4k^2 x(t)\int_0^t x(t)\frac{d^3 x(t)}{dt^3}\, dt + 4k^2\nu^2 x(t) \int_0^t x(t)\frac{dx(t)}{dt}\, dt = 0 \tag{14}$$

The integrals appeared in the equation (14), may be calculated as

$$\int_0^t x(t)\frac{d^3 x(t)}{dt^3}\, dt = x(t)\frac{d^2 x(t)}{dt^2} - \frac{1}{2}\left(\frac{dx(t)}{dt}\right)^2 + C_1, \tag{15}$$

$$\int_0^t x(t)\frac{dx(t)}{dt}\, dt = \frac{1}{2}x^2(t) + C_2, \tag{16}$$

with the constants related to the initial conditions, as

$$C_1 = -x(0)x''(0) + \frac{1}{2}{x'}^2(0), \tag{17}$$

$$C_2 = -\frac{1}{2}x^2(0). \tag{18}$$

In this way, the NDE (14) takes the form

$$\frac{d^4 x(t)}{dt^4} + \left(\nu^2 + \omega^2 + 4k^2 x^2(t)\right)\frac{d^2 x(t)}{dt^2} + (\nu^2\omega^2 - 4\lambda^2\nu\omega + 4k^2 C_1 + 4k^2\nu^2 C_2 - 2k^2\left(\frac{dx(t)}{dt}\right)^2)x(t) + 2k^2\nu^2 x^3(t) = 0. \tag{19}$$

The equation (19) will be solved by the HBM in the next subsection.

### 3.2. The HBM
The HBM is a solution method of the differential equations $F(x, x', x'', \ldots, t) = 0$ in the frequency domain. Starting with the truncated Fourier's series $x_N(t) = \sum_{n=1}^{N}[A_{2n}\cos(n\Omega t) + B_{2n+1}\sin(n\Omega t)]$ as the $N^{\text{th}}$ order harmonic balance ansatz, the HBM procedure tends to a system of algebraic equation including $A_n$, $B_n$ and $\Omega$. Occasionally in a variety of problems, $x_N(t)$ gives reasonably accurate solutions. Besides linear problems, the HBM can be applied to determine both stable and unstable regions of the weak and strong nonlinear differential equations [40]. The HBM is applicable to autonomous and non-autonomous problems, and can thus be used for free, self-excited and externally driven oscillations to calculate the steady-state response of nonlinear differential equations [41].

As we intend to use the HBM to solve the NDE (19), it is indispensable to discuss the convergence of the method. At first, as the HBM is a truncated Fourier series method which convergence has proved [42], the solutions obtained by this method may reach to desired accuracy by choosing suitable truncated

Fourier series. Besides, it has proved that when the exact solution of the problem is known to be analytic, the absolute error of the HBM decreases with an exponential rate [43].

The first order HBM solution of (19) starts by the ansatz

$$x_1(t) = A_1\cos(\Omega_1 t) + B_1 \sin(\Omega_1 t). \tag{20}$$

Hereafter the function $x(t)$ of the equation (19) will be indexed by the order of applied HBM. Tuning the initial condition to have $\frac{dx_1(t)}{dt} = 0$, results in $B_1 = 0$. Considering $\hat{x}(t=0) = \frac{\lambda}{k}(\hat{a} + \hat{a}^\dagger)$, its expectation value in the initial state $|\alpha\rangle$ obtains as $x_{10} = \langle \hat{x}_1(0) \rangle = \langle \alpha | \hat{x}(t) | \alpha \rangle = \frac{2\alpha\lambda}{k}$, in which $\alpha$ is considered as a real parameter, without loss of generality. So $x_1(t) = \frac{2\alpha\lambda}{k}\cos(\Omega_1 t)$ from which we may determine the integration constants as $C_{11} = \frac{4\alpha^2\lambda^2\Omega_1^2}{k^2}$ and $C_{12} = -\frac{2\alpha^2\lambda^2}{k^2}$. Again the first indexes point to the order of HBM. In this way the equation (19) reduces to

$$\frac{d^4 x_1(t)}{dt^4} + (\omega^2 + v^2 + 4k^2 x_1^2(t))\frac{d^2 x_1(t)}{dt^2} + (\omega^2 v^2 - 4\lambda^2 v\omega + 16\alpha^2\lambda^2\Omega_1^2 - 8\alpha^2\lambda^2 v^2 - 2k^2(\frac{dx_1(t)}{dt})^2)x_1(t) + 2k^2 v^2 x_1^3(t) = 0. \tag{21}$$

Following to the first order HBM procedure [38], we set $x_1(t) = A_1\cos(\Omega_1 t)$ in (21) to obtain

$$\cos(\Omega_1 t)\left(\Omega_1^4 + \frac{1}{2}A_1^2 k^2 \Omega_1^2 - v^2 \Omega_1^2 - \omega^2 \Omega_1^2 - 4\lambda^2 v\omega + v^2\omega^2 - \frac{1}{2}A_1^2 k^2 v^2\right) + \frac{1}{2}\cos(3\Omega_1 t)(A_1^2 k^2 v^2 - A_1^2 k^2 \Omega_1^2) = 0, \tag{22}$$

According to the first order HBM, neglecting the terms including $\cos(3\Omega_1 t)$ and equating the coefficient of $\cos(\Omega_1 t)$ to zero, we have

$$\Omega_1^4 + (\frac{1}{2}A_1^2 k^2 - v^2 - \omega^2)\Omega_1^2 + v^2\omega^2 - 4\lambda^2 v\omega - \frac{1}{2}A_1^2 k^2 v^2 = 0. \tag{23}$$

Defining

$$M_1 = \frac{1}{2}A^2 k^2 - v^2 - \omega^2, \tag{24}$$
$$N_1 = -4\lambda^2 v\omega + v^2 \omega^2 - \frac{1}{2}\frac{1}{2}A_1^2 k^2 v^2, \tag{25}$$

the equation (23) takes the form $\Omega_1^4 + M_1 \Omega_1^2 + N_1 = 0$ with the solution

$$\Omega_1 = \frac{1}{\sqrt{2}}(\sqrt{M_1^2 - 4N_1} - M_1)^{\frac{1}{2}}, \tag{26}$$

that is a real frequency for $N_1 < 0$. In this way the solution of (21) by the first order HBM, gives

$$x_1(t) = x_{10}\cos(\frac{t}{\sqrt{2}}(\sqrt{M_1^2 - 4N_1} - M_1)^{\frac{1}{2}}). \tag{27}$$

Using (27), other unknown functions in the system of NDEs (4) to (8), may be calculated as

$$\sigma_{1x}(t) = \sigma_{1x}(0)\cos(\Omega_1 t), \tag{28}$$
$$\sigma_{1y}(t) = \frac{1}{\omega\Omega_1}\sigma_{1x}(0)\sin(\Omega_1 t), \tag{29}$$
$$\sigma_{1z}(t) = \frac{2kx_{10}}{\omega\Omega_1^2}\sigma_{1x}(0)\sin^2(\Omega_1 t) - 1, \tag{30}$$
$$p_1(t) = -mx_{10}\Omega_1 \sin(\Omega_1 t), \tag{31}$$

With $\sigma_{1x}(0) = \frac{mx_{10}}{\hbar k}(\Omega_1^2 - v^2)$. Again, the first indexes in the left hand sides point to the first order HBM.

Second order HBM starts with the ansatz

$$x_2(t) = A_2 \cos(\Omega_2 t) + z_{21} A_2 \cos(3\Omega_2 t). \tag{32}$$

Evaluating the integration constants from (17) and (18) according to (32) and following procedure applied to the first order HBM for the NDE (19), tends to the algebraic equations

$$\Omega_2^4 + M_2 \Omega_2^2 + N_2 = 0, \tag{33}$$
$$az_{21}^3 + bz_{21}^2 + cz_{21} + d = 0, \tag{34}$$

for the coefficients of $\cos(\Omega_2 t)$ and $\cos(3\Omega_2 t)$, with the definitions of $M_2$ and $N_2$ as

$$M_2 = M_1 - 11A_2^2 k^2 z^2 + \frac{53}{2}A_2^2 k^2 z, \tag{35}$$

$$N_2 = N_1 + A_2^2 k^2 \nu^2 z^2 - \frac{5}{2} A_2^2 k^2 \nu^2 z, \tag{36}$$

and

$$a = \frac{9}{2} A_2^2 k^2 \Omega_2^2 - \frac{1}{2} A_2^2 k^2 \nu^2, \tag{37}$$

$$b = 40 A_2^2 k^2 \Omega_2^2 - 4 A_2^2 k^2 \nu^2, \tag{38}$$

$$c = A_2^2 k^2 \nu^2 - 19 A_2^2 k^2 \Omega_2^2 - 4\lambda^2 \nu\omega + \nu^2 \omega^2 - 9\nu^2 \Omega_2^2 - 9\omega^2 \Omega_2^2 + 81\Omega_2^4, \tag{39}$$

$$d = \frac{1}{2} A_2^2 k^2 \nu^2 - \frac{1}{2} A_2^2 k^2 \Omega_2^2 \tag{40}$$

According to the initial conditions, $x_2(t)$ given by (32) must be equal to $\frac{2\alpha\lambda}{k}$ at $t = 0$, so

$$\frac{2\alpha\lambda}{k} = A_2(1 + z_{21}). \tag{41}$$

The equation (33), (34) and (41) may be solved as a coupled system of algebraic equations to obtain $A_2, \Omega_2$ and $z_{21}$ values. The definitions (34) and (35) show that the coefficients $M_1$ and $N_1$ are modified by a systematic trend in the second order HBM. The effects of these changes will be considered numerically besides the third order HBM.

### 3.3. Numerical Results

The numerical analysis of the Rabi model is performed up to the third order HBM for the typical numeric values, [10, 35], given as $\nu = 1$, $\omega = 0.7\nu$, $\lambda = 0.8\nu$, $k = 2.7367 \times 10^{-3}$, $\lambda = 7.7136 \times 10^{-6}$, and $\alpha = 1$. The ion mass considered to be $m = 6.63 \times 10^{-26} kg$ for the calcium and the cavity volume chooses to be $V \sim 10^{-9} m^3$ as an intermediate typical value. The third order harmonic balance method starts by

$$x_3(t) = A_3 \cos(\Omega_3 t) + z_{31} A_3 \cos(3\Omega_3 t) + z_{32} A_3 \cos(5\Omega_3 t), \tag{42}$$

that converts to the second order ansatz when $z_{32} = 0$ and shows the first order one when $z_{31} = z_{32} = 0$. The integration constants $C_1$ and $C_2$ given by (17) and (18), will be changed according to the relevant ansatz for different orders of the HBM. Solution of the problem by third order HBM, means determining the parameters $A_3, \Omega_3, z_{31}$ and $z_{32}$, in such a way that (42) coincides to the initial condition.

The numerical results of the HBMs are given in the table 1.

Table 1. Numerical results of the HBM of orders one, two and three.

| $i$ | $A_i$ $\times 10^{-3}$ | $\Omega_i$ | $z_{i1}$ $\times 10^{-12}$ | $z_{i2}$ $\times 10^{-23}$ |
|---|---|---|---|---|
| 1 | 5.637172 | 0.699999999681 | 0 | 0 |
| 2 | 5.637154 | 0.699999999681 | −4.54 | 0 |
| 3 | 5.637154 | 0.699999999681 | −4.54 | −1.77 |

According to the data given in the table 1, the second and third order harmonics don't modify the frequency of the first harmonic. The relative amplitudes of the second to the first harmonics are of the order $10^{-9}$ for both second and third orders of the HBM. The relative amplitudes of third to the first harmonic in the third order HBM is of the negligible order, $10^{-20}$.

To consider the accuracy of the results over time, the left hand side of the equation (19) is defined as the error functions and plotted for the first and second order harmonic balance solutions in the figure 1.

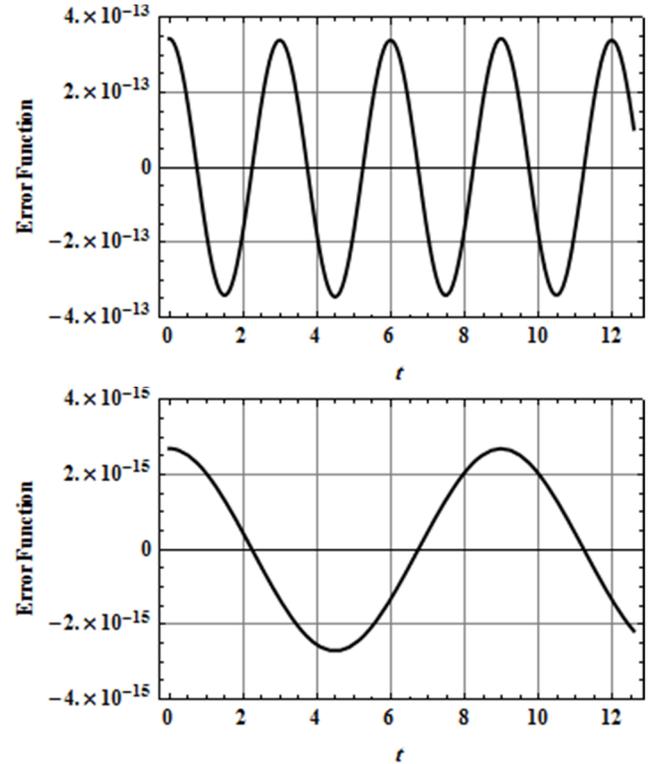

Figure 1. The error functions of the first (Top) and second (Down) order harmonic balance solutions.

The right hand side of (19) equals to zero, so its left hand side must be equal to zero for an exact solution, $x(t)$. In this way, we may consider its left hand side as error function and examine it for any solution. More accurate solution shows smaller deviation of this error function from zero. The figure 1 shows that the maximum deviation from zero is of the order $10^{-13}$ for the solution of the first order HBM while by the second order, the maximum error value reaches to $10^{-15}$ that is reduced by the order of 100 relative the first order's solution.

The numerical solution of the equation (19), $x_n(t)$, and third order harmonic balance solutions are compared through their difference over time as plotted in the figure 2.

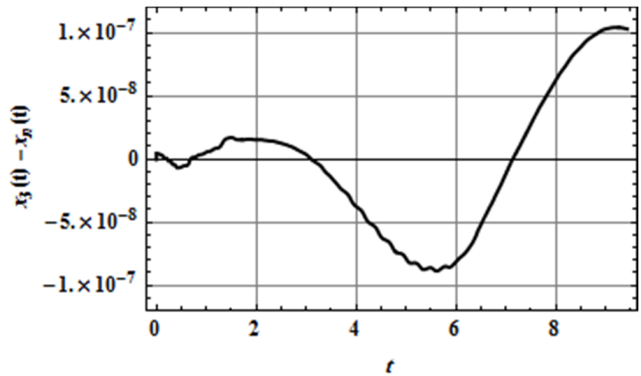

Figure 2. The difference of numerical solution $x_n(t)$, and third order harmonic balance solution of the NDE (19).

The difference of the solutions is negligible for $t < 1$ but it is very large relative to amplitude of the error functions given in the figure 1, implying the HBM is more accurate than the numerical results.

### 3.4. Effects of the RWA
Applying the RWA, the Rabi Hamiltonian (1) in terms of the position and momentum operators takes the form

$$\widehat{H}_{JCM} = \frac{\hbar\omega}{2}\widehat{\sigma}_z + \frac{1}{2m}\hat{p}^2(t) + \frac{1}{2}mv^2\hat{x}^2(t) +$$
$$\lambda\left(\frac{mv\hbar}{2}\right)^{\frac{1}{2}}(\widehat{\sigma}_x(t)\hat{x}(t) - \frac{1}{mv}\widehat{\sigma}_y(t)\hat{p}(t)), \qquad (43)$$

known as the JCM. The Heisenberg equation (3) for the operators appeared in the $\widehat{H}_{JCM}$ gives their time evolutions as the following system of NDEs

$$\frac{d\widehat{\sigma}_x(t)}{dt} = -\omega\widehat{\sigma}_y(t) - s\widehat{\sigma}_z(t)\hat{p}(t), \qquad (44)$$
$$\frac{d\widehat{\sigma}_y(t)}{dt} = \omega\widehat{\sigma}_x(t) - mvs\widehat{\sigma}_z(t)\hat{x}(t), \qquad (45)$$
$$\frac{d\widehat{\sigma}_z(t)}{dt} = mvs\widehat{\sigma}_y(t)\hat{x}(t) + s\widehat{\sigma}_x(t)\hat{p}(t), \qquad (46)$$
$$\frac{d\hat{x}(t)}{dt} = \frac{1}{m}\hat{p}(t) - \frac{\hbar}{2}s\widehat{\sigma}_y(t), \qquad (47)$$
$$\frac{d\hat{p}(t)}{dt} = -mv^2\hat{x}(t) - \frac{mv\hbar}{2}s\widehat{\sigma}_x(t), \qquad (48)$$

in which we used $\widehat{\sigma}_\pm(t) = \frac{1}{2}(\widehat{\sigma}_x(t) \pm i\widehat{\sigma}_y(t))$ and defined $s = \frac{2\lambda}{mv\hbar}\left(\frac{mv\hbar}{2}\right)^{\frac{1}{2}}$.

The system of NDEs (44) to (48), describe the Rabi model under the RWA. The nonlinear terms in the equations (44) to (46) included $\widehat{\sigma}_i(t)\hat{x}(t)$ and $\widehat{\sigma}_i(t)\hat{p}(t)$ factors for $i$ in the $\{x, y, z\}$. As a well localized ion, like a trapped ion, may not have small momentum; the MFT is not applicable to both $\widehat{\sigma}_i(t)\hat{x}(t)$ and $\widehat{\sigma}_i(t)\hat{p}(t)$, simultaneously and so the system of equations (44) to (48) may not be solved as done for the Rabi model.

### 4. Discussion
The MFT has been commonly used to simplify the Dicke model [44] and Tavis-Cummings model [45] that are the multiparticle versions of the Rabi and JCM, respectively. In this context, the MFT is an approximation converting a many body to a single body system [46] in the thermodynamics limit for which the number of particles tends to infinity [47].

The MFT as applied in this work is a different concept by which we rewrite the position operator in terms of the magnitude of fluctuations around its mean, as proved in the subsection 3.1. For small fluctuations, this concept of MFT can be viewed as the "zeroth-order" expansion of the operator in terms of spatial fluctuations around the mean value. This is studied as the dynamic MFT in the quantum field theory for the field operator.

Michael Tavis and Frederick W. Cummings in their seminal paper [45] referred to [5], wherein the JCM has introduced, to explain the validity of their model in the low intensity field conditions and stated that the RWA breakdowns in the high density field. They explained that for the sufficiently weak fields, the counter-rotating terms violate the energy conservation law, when the first order perturbation method applies to solve the problem. According to the proportionality of the position and electric field operators, low field is equivalent to small region in which the ion may be observed as we considered obtaining the approximation given by (27). So the RWA is valid for both of the Dicke and Rabi models. However there are results showing the violation of the RWA in the many body systems. For instance, P. W. Milonni et al. found chaotic conditions in the Dicke model and show that revealing chaos breaks down by applying the RWA [48].

Another problem arising from the RWA when applies to the Rabi model is that the expectation value of the linear equation (47) is explicitly inconsistent to the Ehrenfest's theorem while for the Rabi model, $\frac{dx(t)}{dt} = \frac{1}{m} p(t)$ obtains directly by calculating the expectation value of the equation (7). Inconsistency of the RWA to the Ehrenfest's theorem has been reported through a different analysis for a charged particle irradiated by an external field through the solution of Langevein equation [26].

## 5. Conclusions

The MFT applied to convert the Rabi model to a system of NDEs for the expectation values of the operators In the Heisenberg picture. The model was reduced to a fourth order NDE for position expectation value as a function of time and solved by the HBM. The results are applicable to the Dicke model in similar conditions. Besides the acceptable results of first order HBM, the numerical analysis showed fast reduction of error function in the second order HBM. The HBM may be further considered as a useful and reliable method in the quantum dynamics analysis.

In a similar procedure, the system of NDEs describing the JCM we reduced to a system of NDEs in which simultaneous existence of nonlinear product terms of Pauli matrixes to both position and momentum operators prevented the application of MFT for this model. Besides this inconsistency to the dynamics mean field theory, under the RWA, the equation describing time evolution of the position operator violated the Ehrenfest's theorem, contrary to the Rabi model.